\documentclass[11pt,a4paper]{article}

\usepackage{amsmath}
\usepackage{amsfonts}
\usepackage[parfill]{parskip}
\usepackage{subfig}
\usepackage{graphicx}
\usepackage{sectsty}

\newcommand{\be}{\begin{equation}}
\newcommand{\ee}{\end{equation}}
\newcommand{\ba}{\begin{equation} \begin{aligned}}
\newcommand{\ea}{\end{aligned} \end{equation}}
\newcommand{\ddt}[1]{\frac{\mathrm{d}#1}{\mathrm{d}t}}

\title{\textbf{For principled model fitting in mathematical biology}}
\author{Thomas House\footnote{Warwick Mathematics Institute, University of Warwick,
Coventry, CV4 7AL, UK.}}

\begin{document}

\maketitle

\section*{Introduction}

The mathematical models used to capture features of complex, biological systems
are typically non-linear~\cite{Murray:2002,Murray:2003}, meaning that there are
no generally valid simple relationships between their outputs and the data that
might be used to validate them. This invalidates the assumptions behind
standard statistical methods such as linear regression, and often the methods
used to parameterise biological models from data are \textit{ad hoc}.

In this perspective, I will argue for an approach to model fitting in
mathematical biology that incorporates modern statistical methodology without
losing the insights gained through non-linear dynamic models, and will call
such an approach \textit{principled} model fitting. Principled model fitting
therefore involves defining likelihoods of observing real data on the basis
of models that capture key biological mechanisms.

While few would argue with these general principles, traditionally it has been
considered necessary either to sacrifice likelihood-based model fitting, or
biological realism, for pragmatic reasons.  I argue, however, that while some
level of pragmatism is always necessary, mathematical biologists should be
increasingly able to adopt a principled rather than a pragmatic approach when
fitting models to data.

Of course, the massive and continuing increases in computational power and
data availability play a major role in enabling the possibility of fitting
mechanistic models to data statistically, but mathematical results are at least
as important in my opinion.  First, there is increasing use of stochastic
modelling in biology, and stronger analytical results continue to be proved
about these models, including the ability to relate standard
differential-equation models to underlying stochastic
processes~\cite{AB00,Black:2012}. Secondly, there has been extensive
development of computationally intensive inference algorithms that can in
principle deal with arbitrary likelihood
functions~\cite{Gilks:1995,Brooks:2011}.

In the next few sections, I will develop three points in favour of a principled
approach to model fitting: (1) accurate parameter estimation; (2) uncertainty
quantification; and (3) the role of mechanism. These will be illustrated by
examples using the SIR epidemic model with transmission rate $\beta$ and recovery
rate $\gamma$, technical details of which are collected in the Appendix. The
examples are designed to be simple enough to support the point argued, but
still to exhibit features of real-world problems.

\section{Accurate parameter estimation}

The primary reason for use of principled model fitting is to obtain accurate
estimates of parameters, typically to predict likely future behaviour of a
biological system.  Suppose, for example, we are able to make a full
observation of an infectious disease epidemic, including the onset of and
recovery from disease, as may be possible in small populations such as boarding
schools~\cite{CDSC:1978}. The first example, simplifying such a realistic
scenario, is the general stochastic epidemic~\eqref{gse} with parameters
$N=500$, $\beta = 3$ and $\gamma = 1$. Let us further consider two possible
data sets: (I) observation of the full epidemic; (II) observation until
prevalence of infection is 50 for the first time.

Running Markov chain Monte Carlo (MCMC) with improper priors, using the
tractable likelihood~\eqref{like} for this model, it is possible to
obtain parameter point estimates and 95\% CIs of (I) $\hat{\beta} =
3.02[2.75,3.31]$, $\hat{\gamma}=0.925[0.844,1.01]$ for full data and (II)
$\hat{\beta} = 2.77[2.20,3.53]$, $\hat{\gamma}=0.718[0.419,1.06]$ for early
data only.  As one would hope, the true values sit within the 95\% CIs. Also,
note that while this is a Bayesian procedure, similar results could be obtained
in a frequentist framework, for example by using bootstrapping and numerical
optimisation. 

\begin{figure}
\centering
\subfloat[]{
\scalebox{0.45}{\resizebox{\textwidth}{!}{ \includegraphics{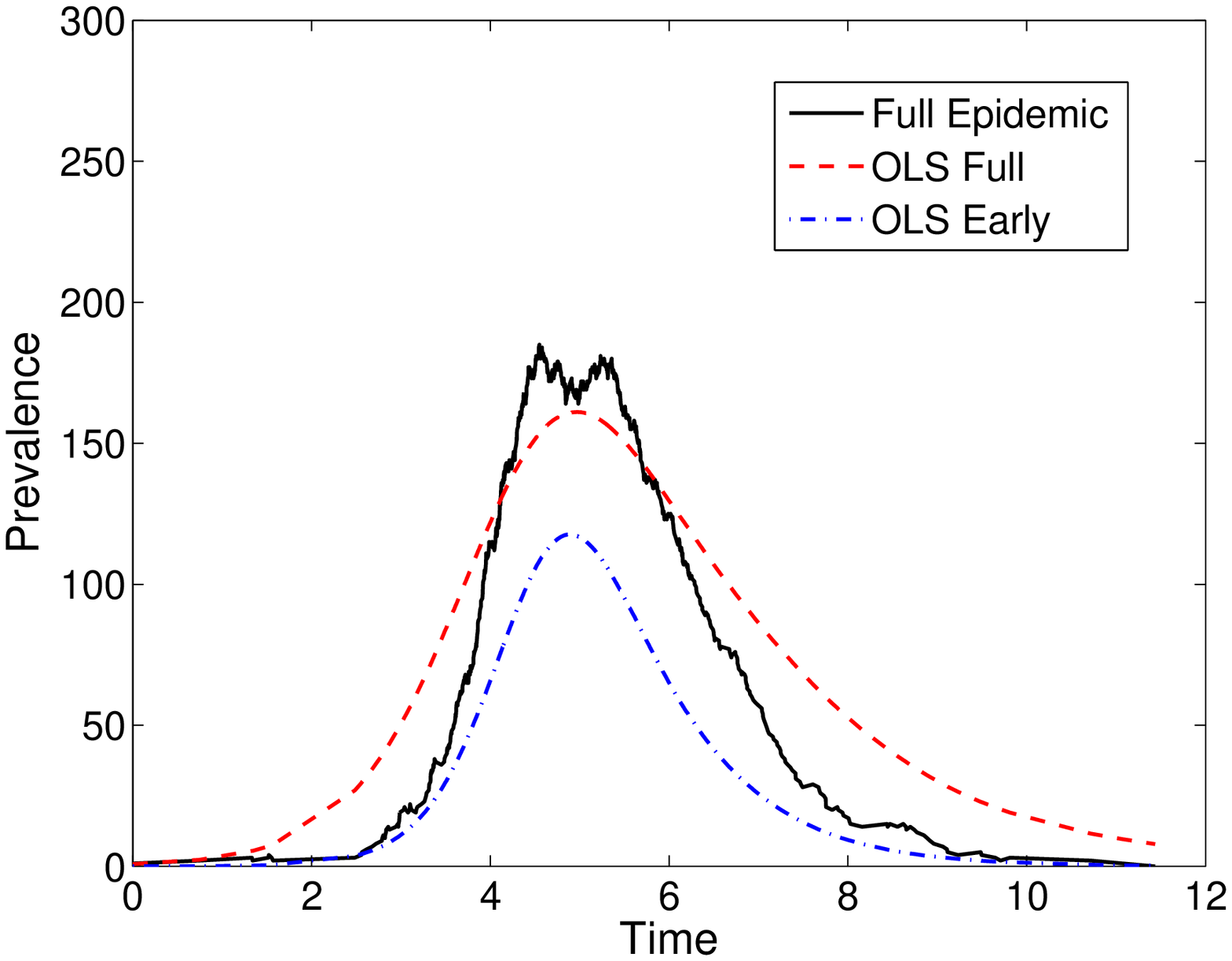} }}}
\quad
\subfloat[]{
\scalebox{0.45}{\resizebox{\textwidth}{!}{ \includegraphics{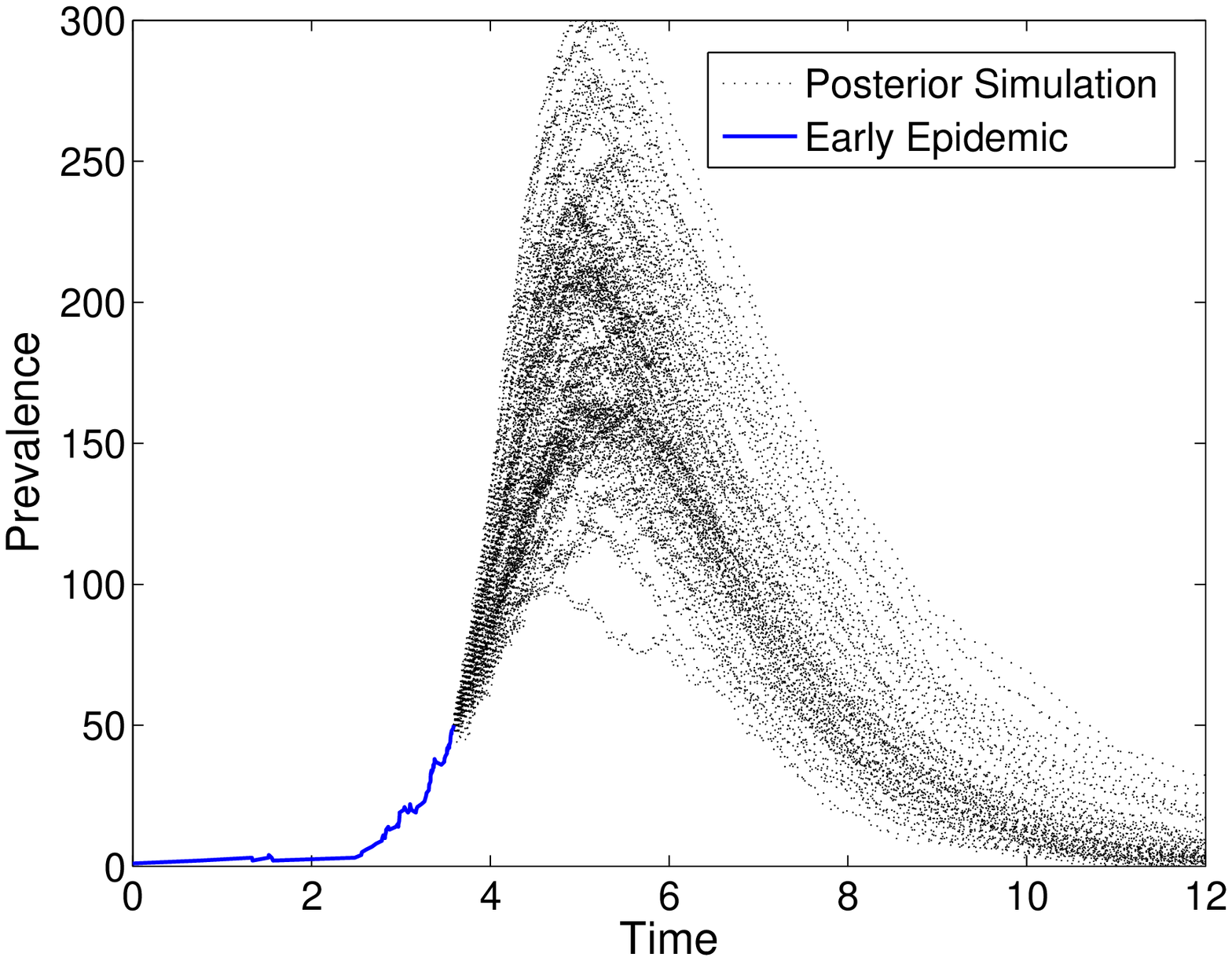} }}}
\caption{Results for fitting to a realisation of the general stochastic
epidemic with $N=500$, $\beta = 3$ and $\gamma = 1$, and $\tau$ defined as the
first time at which $I=50$. (i) Least-squares fitting to the full data (dashed
line) and early data (dash-dotted line). (ii) 100 simulations from independent
samples of the posterior obtained using Bayesian MCMC on the early data.
\label{fitfig}}
\end{figure}

Suppose instead, we fit the standard differential-equation SIR
model~\eqref{sirode} to the data through numerical minimisation of the sum of
squared differences~\eqref{odeols}.  In this case the parameter estimates are
(I) $\hat{\beta} = 2.16$, $\hat{\gamma} = 0.687$ for full data and (II)
$\hat{\beta} = 3.70$, $\hat{\gamma} = 1.48$.  Figure~\ref{fitfig}(i) shows that
neither of these fits is a good description of the data.
Figure~\ref{fitfig}(ii), however, shows that the principled method applied to
early data provides useful predictions for future behaviour of the epidemic.
Crucially, these predictions quantify uncertainty due to finite data and finite
population size, and this is the second argument I wish to make in favour of
principled model fitting.

\section{Uncertainty quantification}

\begin{figure}
\centering
\subfloat[]{
\scalebox{0.45}{\resizebox{\textwidth}{!}{ \includegraphics{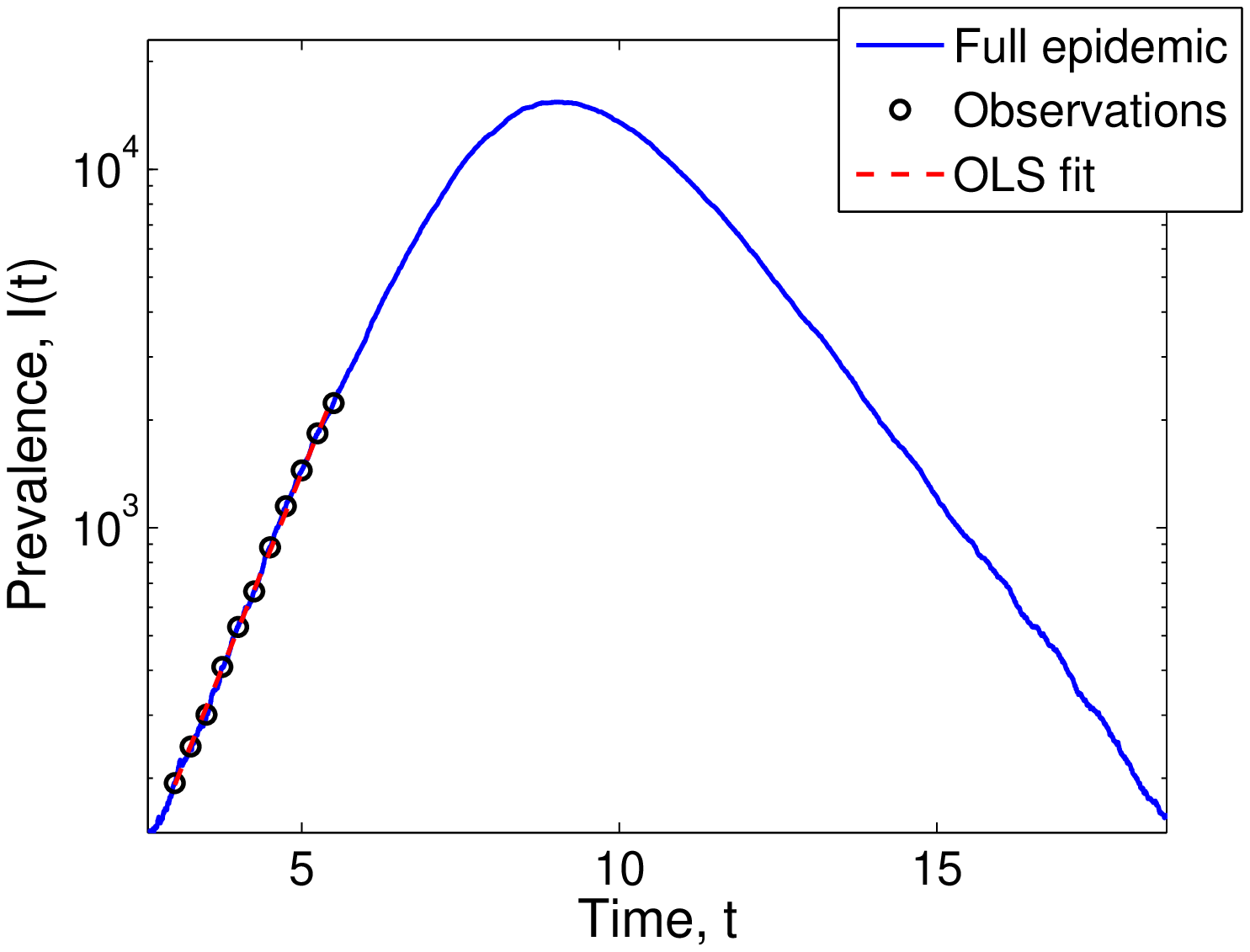} }}}\quad
\subfloat[]{
\scalebox{0.45}{\resizebox{\textwidth}{!}{ \includegraphics{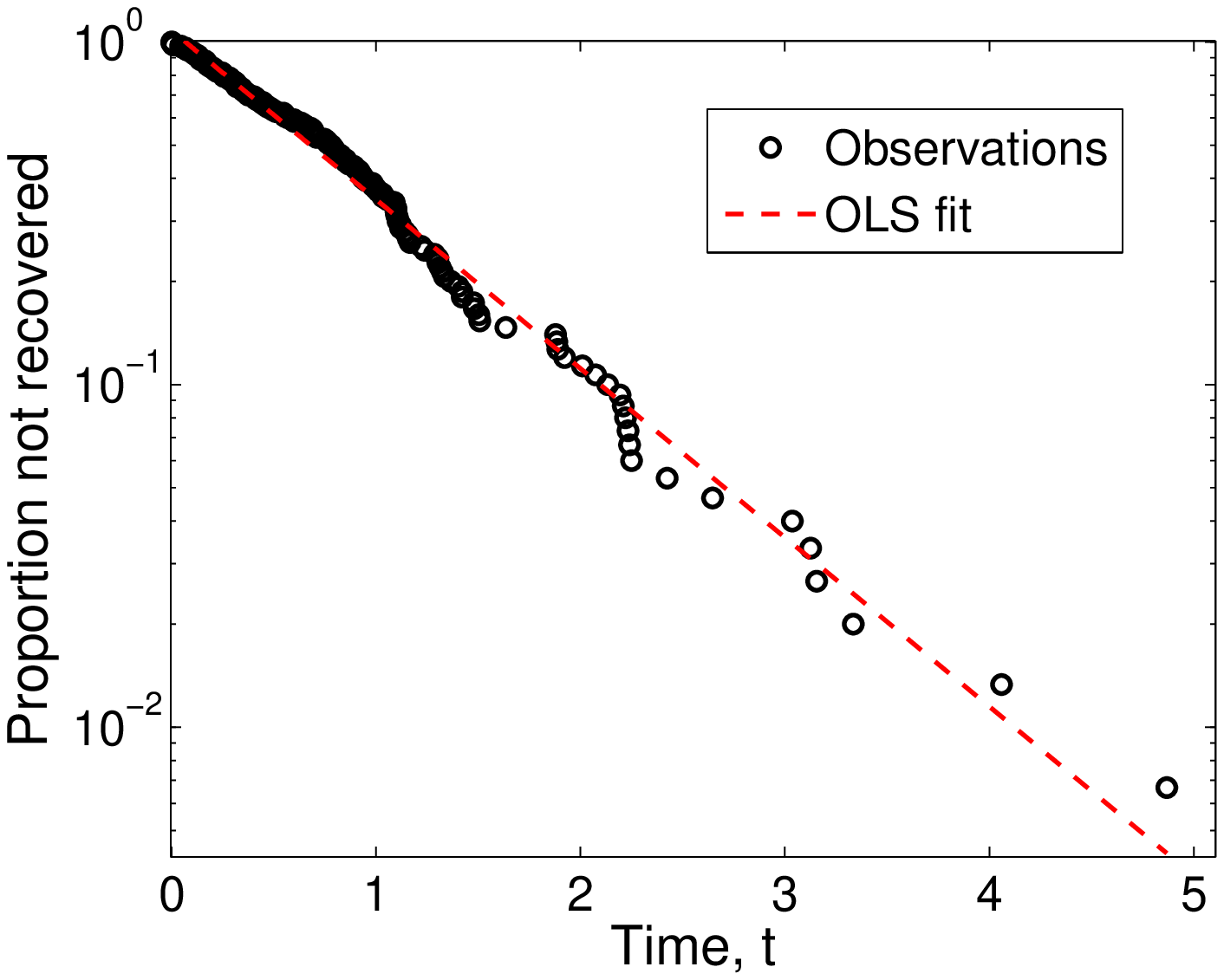} }}}
\\
\subfloat[]{
\scalebox{0.9}{\resizebox{\textwidth}{!}{ \includegraphics{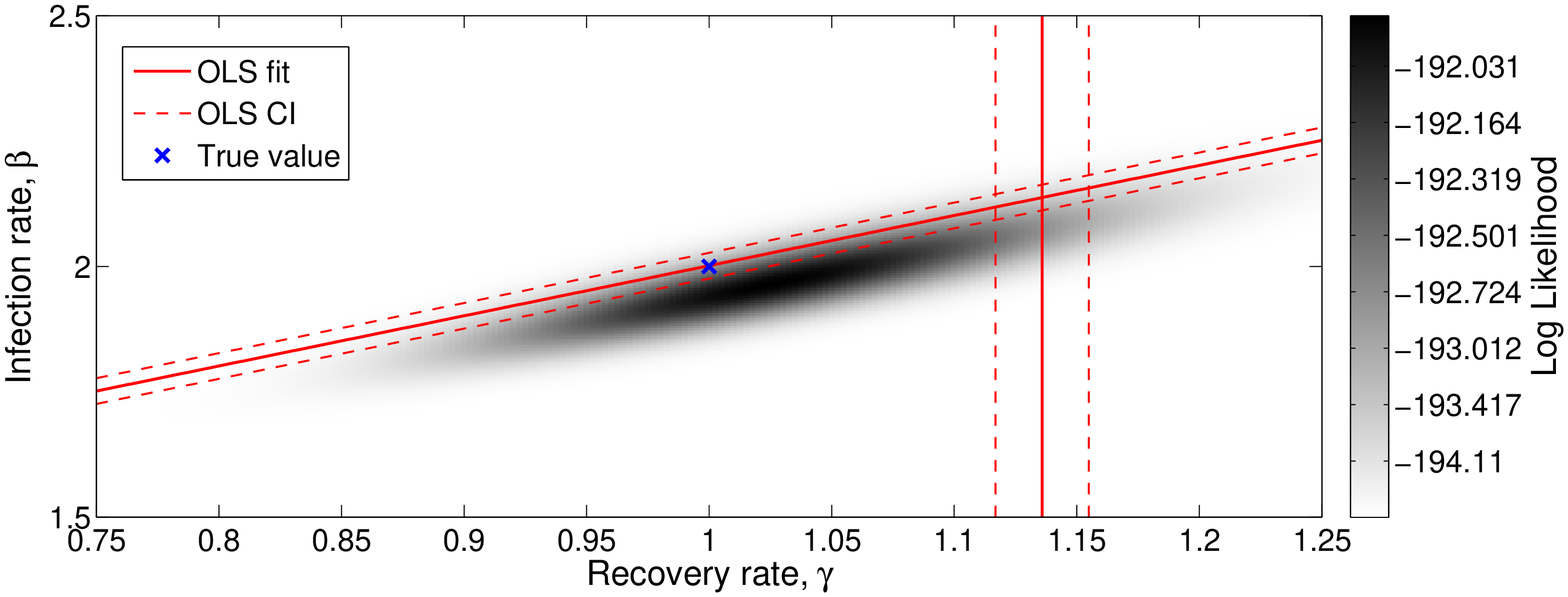} }}}
\caption{
Results for fitting to (i) a realisation of the general stochastic epidemic
with $N=10^5$, $\beta = 2$ and $\gamma = 1$, where prevalence is measured at 11
time points at intervals $\delta = 0.5$ and (ii) $n=150$ observations of times
to recovery from infection. (iii) shows the joint likelihood density, the true
parameter values and the estimates from OLS methods.  \label{lfig}}
\end{figure}

Another important feature of all real data is that it always leaves some
questions unanswered, and quantifying the uncertainties that remain is almost
always the second most important task after the strongest inferences that are
possible have been made. Consider the case where an early epidemic in a large
population is partially observed, so that there are a small number of time
points at which prevalence is known. In this case, the `gold standard' approach
is generally considered to be MCMC with multiple
imputation~\cite{ONeill:1999,Jewell:2009a}, however this remains hard to
implement and so here we use a tractable approximate likelihood~\eqref{gplik}
(thereby illustrating the important point that some level of pragmatism is
often beneficial). Such data is often collected together with more selective
sampling of individual cases to measure the course of infection. We simulate
such a scenario using the model~\eqref{gse} with parameters $N=10^5$, $\beta =
2$ and $\gamma = 1$, taking 11 samples of prevalence early in the epidemic as
shown in Figure~\ref{lfig}(i).  We further sample the recovery times of $n=150$
cases, with survivor function ($1-$the cumulative distribution function) shown
in Figure~\ref{lfig}(ii).

For these sources of data, there are linear relationships that can be used to
obtain ordinary least-squares (OLS) parameter estimates. The first of these is
that early in the epidemic, log$(I(t)) = rt + \mathrm{const.}$, where $r:=
\beta - \gamma$; the second is that the slope of the natural logarithm of the
survival function for recovery is $-\gamma$.  Figure~\ref{lfig}(iii) shows that
while these OLS methods yield sensible estimates of parameters, it is their
confidence intervals that are misleading, since these suggest high confidence
in a region of parameter space far from the true value. In contrast, the true
value \textit{is} in a region of high likelihood density, where the
likelihood is formed from the product of~\eqref{gplik} and the probabilities of
observed recovery times.

\section{The role of mechanism}

\begin{figure}
\centering
\subfloat[]{
\scalebox{0.45}{\resizebox{\textwidth}{!}{ \includegraphics{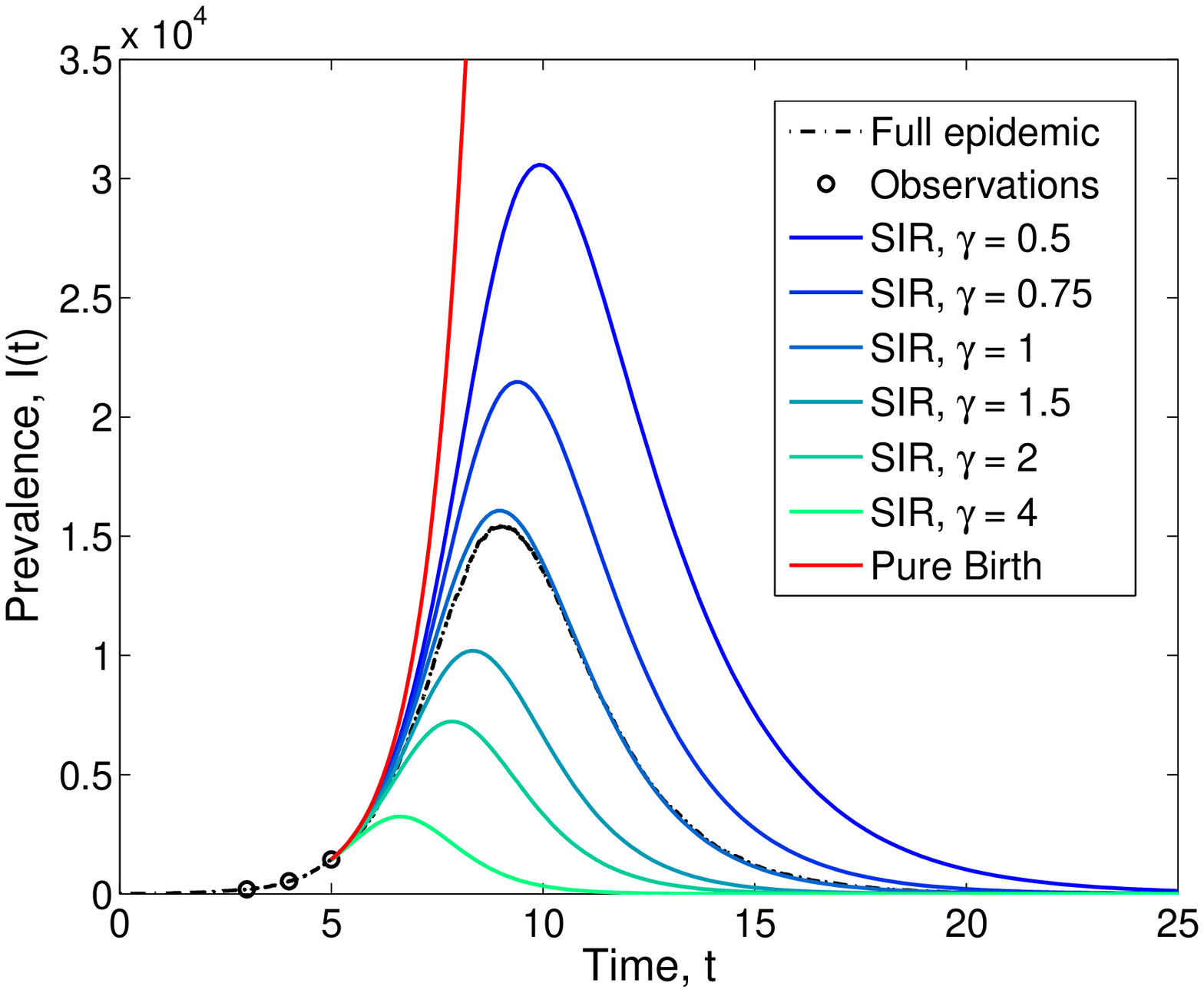} }}}
\quad
\subfloat[]{
\scalebox{0.45}{\resizebox{\textwidth}{!}{ \includegraphics{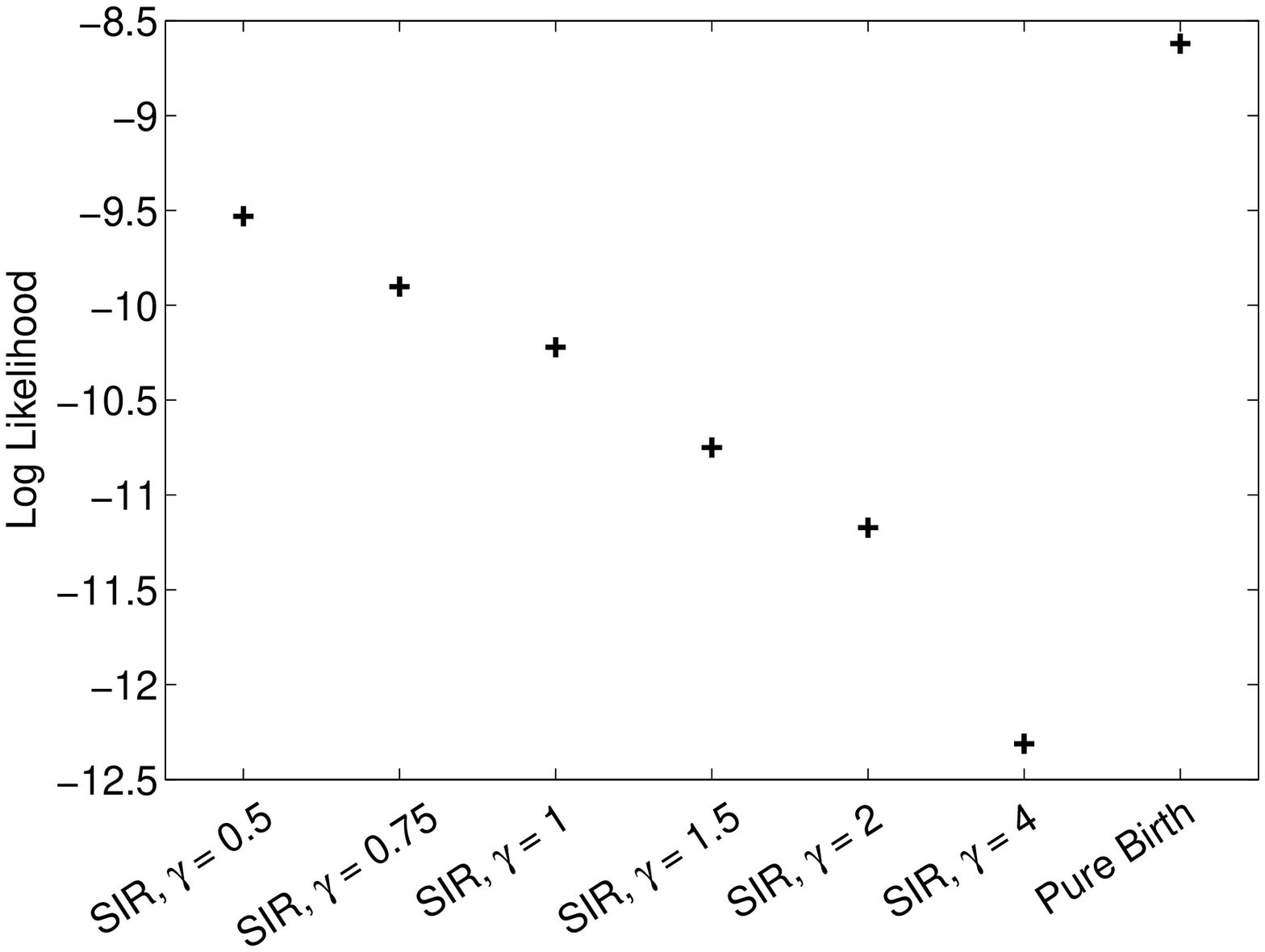} }}}
\caption{
Results for fitting to a realisation of the general stochastic epidemic with
$N=10^5$, $\beta = 2$ and $\gamma = 1$, where prevalence is measured at 11 time
points at intervals $\delta = 0.5$. (i) Predictions for fitted models with
various values of $\gamma$ and a pure birth process, and (ii) comparison of the
likelihoods of fitted models. \label{selfig}}
\end{figure}

The third point I wish to make is that even the most sophisticated statistical
approach is neither a substitute for scientific understanding of the biological
mechanisms behind data, nor an alternative to consideration of appropriate
applied questions relating to prediction or intervention. Suppose we are in a
situation where the only data available are three prevalence estimates early in
the epidemic.  In this case, we have some information about $r = \beta -
\gamma$, but no independent information about $\gamma$. Figure~\ref{selfig}(i)
shows the behaviour of SIR models with different values of $\gamma$, as well as
a pure birth process (i.e.\ $I\rightarrow I+1$ at rate $r$ with no recovery and
no upper limit to the magnitude of $I$). (ii) shows that model likelihood is
maximised for the pure birth process, but scientifically this is not the
correct model and its long-term predictions (unlimited, never-ending growth of
infection) are very far from reality.

I argue that therefore, in data-poor scenarios, the role of modelling remains
to capture key mechanisms and provide predictions that are conditional on the
values of unknown parameters. In fact, the complexity of biological systems
means that typically data will always be inadequate to estimate absolutely
every parameter of interest, meaning that mathematical biologists should
generally be less keen than pure statisticians are to wield Occam's razor.

\section*{Conclusions}

In conclusion, I have argued in favour of principled methods for model fitting
in mathematical biology, which involve both statistical methods such as
likelihood functions, and also the mechanistic models and insights of
traditional mathematical biology.  While biological systems will never be as
predictable or amenable to precision measurement as physical ones, the
development of models that are both predictive (together with appropriate
quantification of uncertainty) and adequate for description of existing
observations is now increasingly realisable. My perspective is that this trend
will continue, and will be an enormously positive development for the field of
mathematical biology.

\appendix

\section{Technical Appendix}

\subsection{Individual-based model and likelihood}

\label{sec:ibm}

The underlying stochastic process is the \textit{general stochastic epidemic}
model~\cite{Bai57}, which consists of two integer-valued non-independent random
variables in continuous time $S(t)$ and $I(t)$, such that $S(t) + I(t) \leq N$,
where the integer $N$ is the population size. This model has two real-valued
parameters, $\lambda$ and $\gamma$, which have dimensions of inverse time and
determine the rates of processes of the Markov chain:
\be
(S,I) \rightarrow (S-1,I+1) \text{ at rate } \lambda S I \text{ ,}\quad
(S,I) \rightarrow (S,I-1) \text{ at rate } \gamma I \text{ .}
\label{gse}
\ee
And so if the current state is $(S,I)$ then the probability densities for the
next event being an infection or recovery after time $t$ are respectively
\be
\rho_1(t) = \lambda S I {\rm e}^{-(\lambda S + \gamma)I t} \text{ ,}\quad
\rho_2(t) = \gamma I {\rm e}^{-(\lambda S + \gamma)I t} \text{ .}
\ee
We will consider the case where observations are a set of times $T$ and events
$\{e(t) \; | \; t \in T \; \& \; e(t) \in\{1,2\}\}$, so that a likelihood
function can be defined as
\be
\mathcal{L}(\beta,\gamma) = \prod_{t\in T} \rho_{e(t)}(t) \text{ .}
\label{like}
\ee

\subsection{Early diffusion limit}

For a population with large size $N$, the early epidemic prevalence $I(t) \ll
N$ converges $I(t) \rightarrow Y(t)$, where $Y(t)$ obeys the stochastic
differential
equation
\be
\ddt{Y} = (\beta - \gamma) Y + \left(\beta^2 + \gamma^2\right)^{1/2} Y \xi \text{ ,} 
\label{sirsde}
\ee
where $\beta:= N\lambda$.  If we make a series of observations $\{y_m\}$ of
infectious prevalence at times $\{t_m\}$, then we can approximate the
likelihood using a Gaussian process:
\ba
\mathcal{L}(\beta,\gamma) & = \prod_m \mathbb{P}[Y(t_{m+1}) =y_{m+1} | Y(t_m)=y_m] \text{ ,}\\
\mathbb{P}[Y(t+\delta) =y' | Y(t)=y] & \approx \mathcal{N}\bigg(y' \bigg| \mu = y {\rm e}^{(\beta - \gamma) \delta} , \;
\sigma = \left(\beta^2 + \gamma^2\right)^{1/2}\mu \delta \bigg)\text{ ,}\\
\mathcal{N}(x | \mu, \sigma) & := \frac{1}{(2\pi)^{1/2} \sigma} {\rm e}^{-(x-\mu)^2/(2\sigma^2)} \text{ .}
\label{gplik}
\ea

\subsection{Deterministic limit}

In the limit of large $N$ (or more strictly $I(t) \gg 1$) the stochastic
process~\eqref{gse} converges on the well-known SIR equations
\be
\ddt{s} = -\beta s i \text{ ,} \quad
\ddt{i} = \beta s i - \gamma i \text{ ,}
\label{sirmf}
\ee
where
\be
s(t):= \frac{1}{N} \mathbb{E}[S(t)]\text{ ,}\quad
i(t):= \frac{1}{N} \mathbb{E}[I(t)]\text{ .} \label{sirode}
\ee
If one approached data of the kind discussed in \S{}\ref{sec:ibm} with the
equations~\eqref{sirmf}, then an alternative to likelihood-based fitting would
be to employ a `least squares' approach, and choose parameters to minimise the
function
\be
\mathcal{E}(\beta, \gamma, i_0) =
 \sum_{t\in T} \left(I(t) - i(t; \beta, \gamma, i_0)\right)^2 \text{ .}
\label{odeols}
\ee

\end{document}